\documentclass{llncs}
\usepackage{amssymb}
\usepackage[pdftex]{graphicx}
\usepackage{amsfonts} 
\usepackage{amsmath} 
\usepackage{latexsym} 
\usepackage{color}
\newcommand{\keywords}[1]{\par\addvspace\baselineskip\noindent{\bf Keywords:}\enspace\ignorespaces#1}
\newcommand{\VS}[1]{}  
\newcommand{\couic}[1]{}

\newcommand{\NLBox}{{\em NLBox}}
\newcommand{\GenNLBox}{{\em GenNLBox}}
\newcommand{\MagicCoins}{{\em MagicCoins}}
\newcommand{\VBox}{{\em VBox}}
\newcommand{\Parity}{{\em Parity}}
\newcommand{\lra}{\longrightarrow}

\newcommand{\ket}[1]{| #1 \rangle}
\newcommand{\bra}[1]{\langle #1 |}

\newcommand{\trace}{\textrm{Tr}}

\newcommand{\range}{\textrm{range}}

\newcommand{\ve}{\varepsilon}

\newcommand{\C}{\mathcal{C}}
\newcommand{\Z}{\mathbb{Z}}

\newcommand{\ie}{{i.e. }}
\newcommand{\cf}{{c.f. }}

\newcommand{\etc}{{etc. }}
\newtheorem{Def}{Definition}
\newtheorem{Th}{Theorem}

\newtheorem{Con}{Conjecture}
\begin{document}
\pagestyle{plain}

\title{Applying causality principles to the axiomatization of probabilistic cellular automata}

\author{Pablo Arrighi\inst{1,2} \and Renan Fargetton\inst{2} \and Vincent Nesme\inst{3} \and Eric Thierry\inst{1}}

\institute{
\'Ecole normale sup\'erieure de Lyon, LIP, 46 all\'ee d'Italie, 69008 Lyon, France\\
\email{\{parrighi,ethierry\}@ens-lyon.fr}
\and
Universit\'e de Grenoble, LIG, 220 rue de la chimie, 38400 SMH, France\\
\email{renan.fargetton@imag.fr}
\and
Universit\"at Potsdam, Karl-Liebknecht-Str. 24/25, 14476 Potsdam, Germany
\email{nesme@qipc.org}
}

\maketitle
\thispagestyle{empty}

\begin{abstract} 
Cellular automata (CA) consist of an array of identical cells, each of which may take one of a finite number of possible states. The entire array evolves in discrete time steps by iterating a global evolution $G$. Further, this global evolution $G$ is required  to be shift-invariant (it acts the same everywhere) and causal (information cannot be transmitted faster than some fixed number of cells per time step). At least in the classical \cite{Hedlund}, reversible \cite{KariCircuit} and quantum cases \cite{ArrighiUCAUSAL}, these two top-down axiomatic conditions are sufficient to entail more bottom-up, operational descriptions of $G$. We investigate whether the same is true in the probabilistic case.
\end{abstract}

\section{Introduction}

Due to its built-in symmetries, CA constitute a clearly physics-like model of computation \cite{MargolusPhysics}. They model spatially distributed computation in space as we know it \cite{ToffoliMargolusModelling,MooreFiring}, and therefore they constitute a framework for studying and proving properties about such systems -- by far the most established. Conceived amongst theoretical physicists such as Ulam and Von Neumann \cite{Neumann}, CA were soon considered as a possible model for particle physics, fluids, and the related differential equations. There are numerous results on this approach, often under the name of Lattice-gas cellular automata \cite{Zuse,Wolf-Gladrow,ChopardDroz,RothmanZaleski}. More generally, CA are one of the main theoretical tools of `complex sciences', where one studies the emergence of complex, global behaviours as arising from simple local interactions. There, CA have proved useful for modelling an incredible variety of things ranging traffic jams \cite{NagelSchreckenberg} to demographics and regional development or consumption \cite{Bruun,WhiteEngelen}.\\
Each of this variety of contexts brings its own set of reasons to study probabilistic extensions of CA. This trend of work has already started, and generated fascinating questions. For example, are there CA which can defend themselves against an everywhere present noise? Or is it the case that any initial configurations will ultimately be entirely erased, \ie that such CA have only one limit distribution? In \cite{Toom,TVSMKP} it was shown, via an extremely convoluted argument which many would like to simplify \cite{RST,Mairesse,Fates}, that there exists CA which are resistant to this noise model. However, perhaps it would help to address these issues to have a robust definition of Probabilistic CA (PCA).  
\begin{figure}[h]
\centering
\includegraphics[scale=1.0, clip=true, trim=0.5cm 0.5cm 0.5cm 0.3cm]{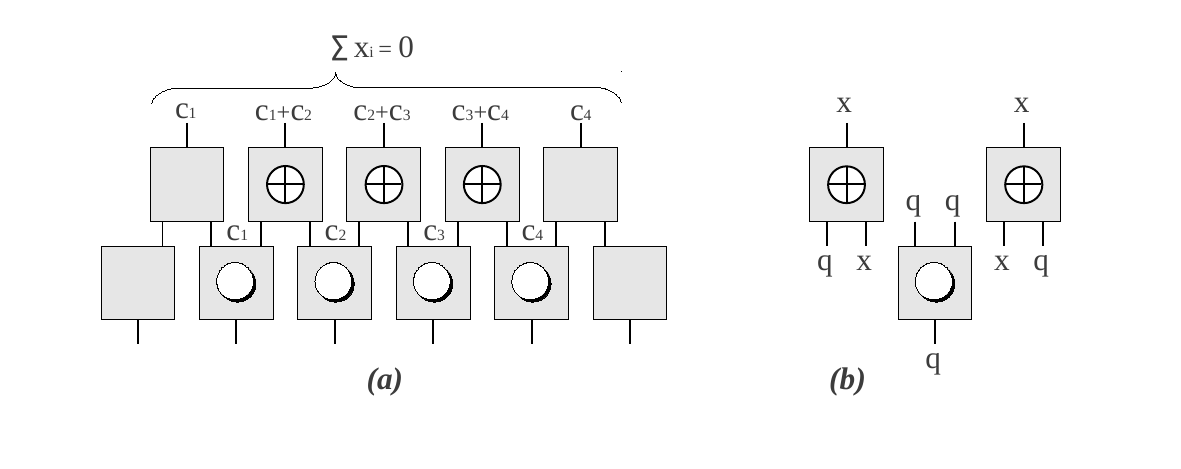}
\caption{
{\bf \emph{(a)}} \Parity.\label{fig:parity} {\small Time flows upwards. Wires are bits. Boxes are (randomized) gates, i.e. stochastic matrices. Boxes marked with a circle are coin tosses. Boxes marked $\oplus$ perform addition modulo two. Input cells $0\ldots n$ are ignored. Output random variables $(X_i)_{i=1\ldots n}$ have the property that for strict subset $I$ of $1\ldots n$, $X_I$ is uniformly distributed, yet their global parity is always $0$.}
{\bf \emph{(b)}} {\small As such \Parity\ is a localized stochastic map, but it can be modified into a translation-invariant stochastic map over entire configurations of the form $\ldots qq01001qq\ldots$, by ensuring that for any $x\in\{q,0,1\}$, $qx\mapsto x$, $xq\mapsto x$, $qq\mapsto q$, and hence making the coin tosses conditional upon their input cells not being $q$. \Parity\ does not pertain to the class of Standard-PCA, but it ought to be considered a valid PCA.}}
\end{figure}
All of these papers begin with a definition of PCA, but these definitions are all variations of the same concept.  Namely, these are stochastic maps having the property that they decompose in two phases: first, a classical CA is applied, and second, a model of noise is applied (a stochastic matrix is applied homogeneously on each individual cell). We refer to this class of stochastic maps as Standard-PCA and now explain its several drawbacks.\\
First, this class is incomplete. There are many stochastic maps which ought to be called PCA but are not Standard-PCA, not even with increased cell or neighbourhood sizes. \Parity\ is a concrete example of this, see Figure \ref{fig:parity}. Indeed, \Parity\ is translation-invariant, implementable by local means, but it may generate statistical correlations between any two adjacent regions of cells, which is never the case of Standard-PCA. Of course \Parity\ can be implemented as the square (i.e. two time-steps) of a Standard-PCA, which points out to a second, even worse problem with this class: the composition of two Standard-PCA is not necessarily a Standard-PCA. This is counter-intuitive for a notion of PCA. Thirdly, the intuition behind this class is simple but ad-hoc; \ie not founded upon some meaningful high-level principles.\\
This paper aims to define PCA in a more robust manner, which is a long-standing problem. Criteria for a good axiomatic definition include: being composable, and being based on high-level principles, while entailing an operational description (i.e. implementation by local mechanisms). 

Why not start right-away from the latter, operational description? If not entailed by well-agreed principles, the operational description may be incomplete and ad-hoc, as with Standard-PCA. Moreover, axiomatic definitions tend to have a practical interest as simple characterizations of the operational descriptions.\\
In fact a great deal of foundational mathematical results work that way. When an axiomatic definition is given an operational description, we speak of a `structure theorem' or a `representation theorem' (e.g. spectral decomposition of a unitary operator as $\sum e^{i\theta_i} \ket{\psi_i}\bra{\psi_i}$). When an operational description is given an axiomatic definition, we talk of a `characterization'
(e.g. Hedlund's characterization of a CA as the set of shift-invariant continuous functions in the Cantor topology).

Why not generalize Hedlund's characterization of CA to PCA? We will begin by investigating precisely that route and show that continuity arguments fail to characterize PCA (Section \ref{sec:continuity}), as they do not forbid spontaneous generation of statistical correlation in arbitrarily distant places (Section \ref{sec:correlations}).  Counter-examples are necessarily non-signalling, non-local stochastic maps. That such objects can exist is now a well-known fact in quantum theory, where non-signalling non-locality arises from entanglement in Bell's test experiment \cite{Bell}. Recently their study has been abstracted away from quantum theory, for instance via the rather topical \NLBox\  \cite{Beckman,PopescuRohrlich,BHK}.

This points out the weakness of the non-signalling condition in the probabilistic/stochastic setting, which is a well-known issue in foundations of physics. In this context, more robust causality principles have been considered.  In fact, Bell's test experiments are motivated by a `local causality' , of which there exist several variants, all of them stemming from Reichenbach's `principle of common cause'. Now, since PCA are nothing but simplifications of those spaces used in physics, this principle of common cause, if it is at all robust, should therefore lead to the axiomatization of PCA. We investigate this intriguing question (Section \ref{sec:commoncause}) and answer it in an unexpected fashion (Section \ref{sec:conclusion}). First we formalize the problem.

\section{Problem statement}\label{sec:problem}

\begin{Def}[Configurations]
A \emph{finite (unbounded) configuration} $c$ over a finite alphabet $\Sigma$ is a function $c:
\Z \longrightarrow \Sigma$, with $i\longmapsto
c(i)=c_i$, such that there exists a (possibly empty)
finite set $I$ satisfying $i\notin I\Rightarrow c_i=q$, where $q$ is a distinguished \emph{quiescent} state of $\Sigma$. We denote $\C$, this set of finite configurations; it is countable. 
\end{Def}
Countability is not crucial to the arguments developed in this paper, but it certainly makes the following definitions much easier:
\begin{Def}[Random variables, states]
Random variables denoted by $X$ range over $\C$, \ie entire configurations.
Random variables denoted by $X_i$ range over $\Sigma$, \ie cells.
Random variables denoted by $X_I$\ range over $I\to\Sigma$, \ie sets of cells.
Random variables corresponding to time $t$, are denoted $X^t$, $X^t_I$.
The \emph{state} of the random variable $Y$, denoted $\rho_Y$, is the functions from $\range(Y)$ to $[0,1]$ such that $\rho_Y(y)=Pr(Y=y)$, with $Pr(Y=y)$ the probability mass function of $Y$ at $y$, \ie they denote the law of distribution. For convenience in the particular case when $Y$ is fully determined on $y$, i.e. such that $\rho_Y(x)=\delta_{xy}$, the state $\rho_Y$ will be written $\tilde{y}$. Moreover for convenience $\rho_{X^t_I}$ will be denoted $\rho^t_I$, and referred to as the \emph{state of cells $I$ at time $t$}. Whenever $I$ is finite, we can make $\rho^t_I$ explicit as the vector $\big(\rho^t_I(w)\big)_{w\in \Sigma^I}$ with the ${|\Sigma|}^{|I|}$ entries listed in the lexicographic order.  
\end{Def}
\begin{Def}[Stochastic maps]
Consider $X$ and $X'$ two random variables over the same range, and their corresponding states. We define the state $p\rho_X+(1-p)\rho_{X'}$ so that $(p\rho_X+(1-p)\rho_{X'})(y)=p\rho_X(y)+(1-p)\rho_{X'}(y)$. Consider $S$ a function from states $(\range(X)\to[0,1])$ to states $(\range(Y)\to[0,1])$. Then $S$ is a {\em stochastic map over the range of $X$} if and only if it is linear, i.e. $S(p\rho_X+(1-p)\rho_{X'})=pS\rho_X+(1-p)S\rho_{X'}$. Whenever $\range(X)$ and $\range(Y)$ are finite, we can make $S$ explicit as the stochastic matrix $S=\big(S\tilde{x}(y)\big)_{y\in\range(Y),x\in\range(X)}$. Notice that each column is the law of distribution $S\tilde{x}$ for some state $\tilde{x}$, hence its entries are in $[0,1]$ and sum to one (left stochasticity). We assume that the random variables over configurations $(X^t)_{t\in\mathbb{N}}$ follow a stochastic process, i.e. that they form a Markov chain $Pr(X^{n+1}=x^{n+1}|X^n=x^n)=Pr(X^{n+1}=x^{n+1}|X^n=x^n,\ldots,X^0=x^0)$ and obey the recurrence relation $\rho^t=G^t\rho$ for some stochastic map $G$ over configurations. 
\end{Def}
Our problem is to determine what it means for a stochastic map over configurations $G$ to be {\em causal}, meaning that arbitrarily remote regions $I$ and $J$ do not influence each other {\em by any means}. Then PCA will just be causal, shift-invariant stochastic maps.

\section{Continuity}\label{sec:continuity}
\begin{figure}[h]
\centering
\includegraphics[scale=1.0, clip=true, trim=0.0cm 2.3cm 0.0cm 1.0cm]{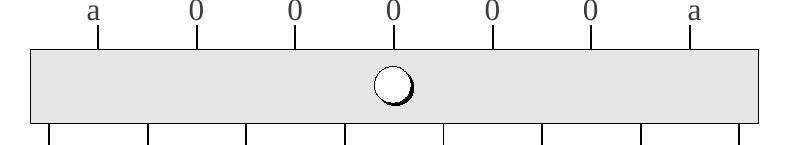}
\caption{\MagicCoins.\label{fig:magiccoins} {\small Inputs are ignored. Hence output random variables $(X_i)_{i=1\ldots n}$ have the property that for any output $x$, the state (a.k.a the law of distribution) $\rho_{x}$ is independent of the inputs. Yet output random variables $X_1$ and $X_{n}$ are both uniformly distributed and maximally correlated. As such \MagicCoins\ is a localized stochastic map, but can be modified into a translation-invariant stochastic map over entire configurations using the method described for \Parity. \MagicCoins\ is non-signalling, but it must not be considered a valid PCA: it is not non-correlating.}}
\end{figure}

In the deterministic case CA were axiomatized by the celebrated Curtis-Lyndon-Hedlund Theorem\cite{Hedlund}, which we now recall. 
First the space of configurations is endowed with a metric:
\begin{Def}[Metric] The function $d(.,.): \mathcal{C}\times\mathcal{C}\lra\mathbb{R}^+$ such that 
$d(c,c')=0$ if $c=c'$ and $d(c,c')=1/2^k$ with $k=\min\{i\in\mathbb{N}\;|\; c_{-i\ldots i}\neq c'_{-i\ldots i} \}$ is a metric. For $c,c'\in\mathcal{C}$ and $\ve>0$ we have (with
$n=\lfloor \log_2(\ve)\rfloor$): 
$d(c,c')<\ve \Leftrightarrow c_{-n\ldots n}=c'_{-n\ldots n}.$
\end{Def}
\begin{Def}[(Uniform) continuity]
A function $F:\mathcal{C}\lra \mathcal{C}$
is continuous if and only if for all $c\in\mathcal{C}$ and $\ve>0$, there
exists $\eta>0$ such that for all $c'\in\mathcal{C}$:
$d(c,c')<\eta\;\Rightarrow\; d(F(c),F(c'))<\ve.$
A function $F:\mathcal{C}\lra \mathcal{C}$
is uniformly continuous if and only if for all $\ve>0$, there
exists $\eta>0$ such that for all $c,c'\in\mathcal{C}$:
$d(c,c')<\eta\;\Rightarrow\; d(F(c),F(c'))<\ve.$
\end{Def}
In other words a function $F:\mathcal{C}\lra \mathcal{C}$
is uniformly continuous if and only if for all $n\in\mathbb{N}$, there exists $m\in\mathbb{N}$ such that for all $c,c'\in\mathcal{C}_\infty$, $c_{-m\ldots m}=c'_{-m\ldots m}$ implies $F(c)_{-n\ldots n}=F(c')_{-n\ldots n}$. Notice that rephrased in this manner, uniform continuity is a synonym for non-signalling, i.e. the fact that information does not propagate faster than a fixed speed bound. Continuity on the other hand expresses a somewhat strange form of relaxed non-signalling, where information does not propagate faster than a certain speed bound, but this speed bound depends upon the input. However, it so happens that the two notions coincide for compact spaces. Moreover, classical CA are easily defined upon infinite configurations $\mathcal{C}_\infty:\mathbb{Z}\to\Sigma$, for which the same $d(.,.)$ happens to be a compact metric. This yields: 
\begin{Th}[Curtis, Lyndon, Hedlund]\label{th:hedlundinf}~\\
A function $F:\mathcal{C}_\infty\lra \mathcal{C}_\infty$ 
is continuous and shift-invariant if and only if
it is the global evolution of a cellular automaton over $\mathcal{C}_\infty$. 
\end{Th}
In other words this theorem just states that CA are exactly the non-signalling, shift-invariant functions. But instead of having to call them `non-signalling' (a.k.a `uniformly continuous'), it only needs to call them `continuous', due to the peculiarities of $\C_\infty$. However for the finite, yet unbounded configurations $\C$, the metric $d(.,.)$ is not compact. In this case we must assume the stronger, uniform continuity for the theorem to work. Generally speaking, it is rather difficult to find a relevant compact metric for probabilistic extensions of $\C_\infty$ --- and not worth the effort for the sole purpose of axiomatizing PCA. Indeed, let us directly assume the probabilistic counterpart of non-signalling (a.k.a uniform continuity for some extended metric):
\begin{Def}[Non-signalling] A stochastic map over configurations $G$ is non-signalling if and only if for any $\rho,\rho'$ two states over configurations, and for any cell $i$, we have:
$$\rho_{i-1,i}=\rho'_{i-1,i}\quad\Rightarrow\quad (G\rho)_i=(G\rho')_i.$$ 
\end{Def}
For example, \Parity\ (see Figure \ref{fig:parity}) is non-signalling by construction. Is it reasonable to say, \`a la Hedlund, that PCA are the non-signalling, shift-invariant stochastic maps? Surprisingly, this is not the case. Imagine that Alice in Paris (cell $0$) tosses a fair coin, whilst Bob in New York (cell $n+1$) tosses another. Imagine that the two coins are magically correlated, \ie it so happens that they always yield the same result. Such a stochastic map is clearly not implementable by local mechanisms: we definitely need some amount of (prior) communication between Alice and Bob in order to make it happen. Yet it can be, as in \MagicCoins\ (see Figure \ref{fig:magiccoins}), perfectly non-signalling. While the setup cannot be used to communicate `information' between distant places, it can be used to create spontaneous `correlations' between them. We must forbid this from happening. In this respect, assuming only (non-uniform) continuity is the wrong direction to take.

\section{Avoiding spontaneous correlations}\label{sec:correlations}

\begin{figure}[h]
\centering
\includegraphics[scale=1.0, clip=true, trim=0.0cm 0.8cm 0.5cm 0.1cm]{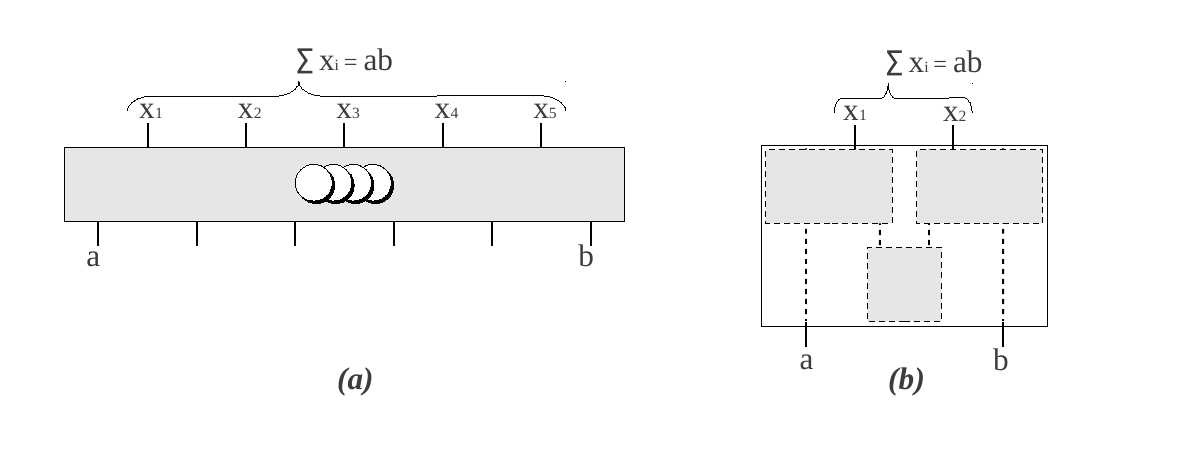}
\caption{{\bf \emph{(a)}} \GenNLBox.\label{fig:AllNLBox} {\small Input cells $1\ldots n-1$ are ignored. Output random variables $(X_i)_{i=1\ldots n}$ have the property that for strict subset $I$ of $1\ldots n$, $X_I$ is uniformly distributed, yet the global parity is always $ab$. As such \GenNLBox\ is a localized stochastic map, but can be modified into a translation-invariant stochastic map over entire configurations using the method described for \Parity. \GenNLBox\ is both non-signalling and non-correlating, but it must not be considered a valid PCA: it is not $V$-causal.}
{\bf \emph{(b)}} \NLBox . {\small The output random variables $X_1$ and $X_2$ are uniformly distributed, but their parity is always equal $ab$. No circuit of the displayed dotted form can meet these specifications \cite{Bell,WernerBell}.} }
\end{figure}

From the previous discussion we are obliged to conclude that the formalization of a robust notion of the causality of a dynamics is indeed a non-trivial matter in a probabilistic setting. From the \MagicCoins\ example, we draw the conclusion that such a notion must forbid the creation of spontaneous correlations between distant places. The following definition clarifies what is meant by (non-)correlation between subsystems.
\begin{Def}[Independence, tensor, trace]
Let $I$ and $J$ be disjoint. Stating that $X_I$ and $X_J$ are independent is equivalent to stating that for any $uv$ in $(I\cup J\to\Sigma)$, $\rho_{I\cup J}(uv)=\rho_{I}(u)\rho_{J}(v)$, with $u$ (resp. $v$) the restriction of $uv$ to $I$ (resp. $J$). In this case we write $\rho_{I\cup J}=\rho_{I}\otimes\rho_{J}$. This notation is justified because whenever $I$ and $J$ are finite, we have that the law of distribution $(\rho_{I\cup J}(uv))$ equals $(\rho_{I}(u)\rho_{J}(v))$ which is the definition of $(\rho_{I}(u))\otimes(\rho_{J}(v))$, where $\otimes$ is the Kronecker/tensor product. Whether or not $X_I$ and $X_J$ are independent, we can always recover $\rho_I$ as the marginal of $\rho_{I\cup J}$ by averaging over every possible $v$, an operation which we denote $\trace_J$ and call the trace-out/marginal-out operation. Namely we have that $\rho_I=\trace_J(\rho_{I\cup J})$, with $\trace_J(\rho_{I\cup J})(u)=\sum_{v}\rho_{I\cup J}(uv)$. 
\end{Def}
A way to forbid spontaneous correlations is to require that, after one-time step of the global evolution $G$ applied upon any initially fully determined configuration $\tilde{c}$, and for any two distant regions $I=-\infty\ldots x-1$ and $J=x+1\ldots\infty$, the output $\rho=G\tilde{c}$ be such that $\rho_{I\cup J}=\rho_{I}\otimes\rho_{J}$. In other words remote regions remain independent. This formulation is somewhat cumbersome, because it seeks to capture in the vocabulary of `states' a property which really belongs to their `dynamics'. The following definition clarifies what it means for a stochastic map to be localized upon a subsystem.
\begin{Def}[Extension, localization, tensor]
Let $I$ and $J$ be disjoint. Consider a stochastic map $S$ over $I$, i.e. from states $((I\to\Sigma)\to[0,1])$ to themselves. Then $S$ can be trivially extended into a stochastic map $S\otimes Id$ over $I\cup J$, i.e. from states $((I\cup J\to\Sigma)\to[0,1])$ to themselves, and such that $(S\otimes Id)\rho_{I}\otimes\rho_{J}=(S\rho_{I})\otimes\rho_{J}$. Whenever $I$ and $J$ are finite, this extends the stochastic matrix $\big(S\tilde{u}(u')\big)$ into $\big(S\tilde{u}(u')\delta_{vv'}\big)$ of $(S\otimes Id)$. We say that some stochastic map over $I\cup J$ is {\em localized upon $I$} precisely if it arises in such away, i.e. as the trivial extension of some stochastic map $S$ over $I$. Moreover if $T$ is over $J$, then $(S\otimes T)=(Id\otimes T)(S\otimes Id)=(S\otimes Id)(Id\otimes T)$. This notation is justified because whenever $I$ and $J$ are finite, we have that the resulting stochastic matrix is $\big(S\tilde{u}(u')T\tilde{v}(v')\big)$ which equals $\big(S\tilde{u}(u')\big)\otimes\big(T\tilde{v}(v')\big)$, the Kronecker/tensor product of both stochastic matrices.
\end{Def}
We can then forbid spontaneous correlations directly in terms of dynamics:
\begin{Def}[non-correlation] A stochastic map over configurations $G$ is non-correlating if and only if for any output cell $i$, there exist stochastic maps $A, B$ acting over input cells $-\infty\ldots x-1$ and $x\ldots +\infty$ respectively, such that: $\quad\trace_x\circ G= A\otimes B.$
\end{Def}
For example, \Parity\ (see Figure \ref{fig:parity}) is non-correlating by construction. Now, is it reasonable to say that PCA are the non-correlating, shift-invariant stochastic maps? Amazingly, this is not the case. Indeed, consider a small variation of \Parity, which we call \GenNLBox\ and define in Figure \ref{fig:AllNLBox}. Such a stochastic map is clearly not implementable by local mechanisms: we definitely need some amount of communication between Alice and Bob in order to make it happen. It suffices to notice that \GenNLBox\ is in fact a generalization of the `non-local box', which we recover for $n=2$, see Figure \ref{fig:AllNLBox}\emph{(a)}. But then, the non-local box owes its name precisely to the fact that it is not implementable by local mechanisms. Formal proofs of this assertion can be found in the literature \cite{Beckman,PopescuRohrlich} and rely on the fact that the \NLBox\  (maximally) violates the Bell inequalities \cite{Bell,WernerBell}. Yet \GenNLBox\ (see Figure \ref{fig:AllNLBox} \emph{(a)}), was perfectly non-correlating. Hence, whilst the set-up cannot be used to communicate `information' between distant places (it is non-signalling), nor to create spontaneous `correlations' between distant places (it is non-correlating), we still must forbid it from happening!
\section{Common cause principle}\label{sec:commoncause}

\begin{figure}[h]
\centering
\includegraphics[scale=0.95, clip=true, trim=0.2cm 1.5cm 0.2cm 0.5cm]{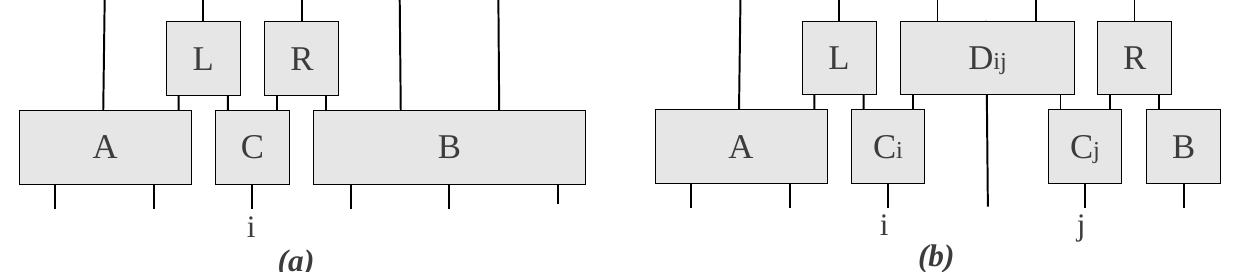}
\caption{\label{fig:VCausal} 
{\bf \emph{(a)}} {\small $G$ is $V$-causal if it can be put in this form, for each $i$.}  
{\bf \emph{(b)}} {\small A strengthened condition: $VV$-causality.}
}
\end{figure}

\begin{figure}[h]
\centering
\includegraphics[scale=1.0, clip=true, trim=0.5cm 0.8cm 0.5cm 0.8cm]{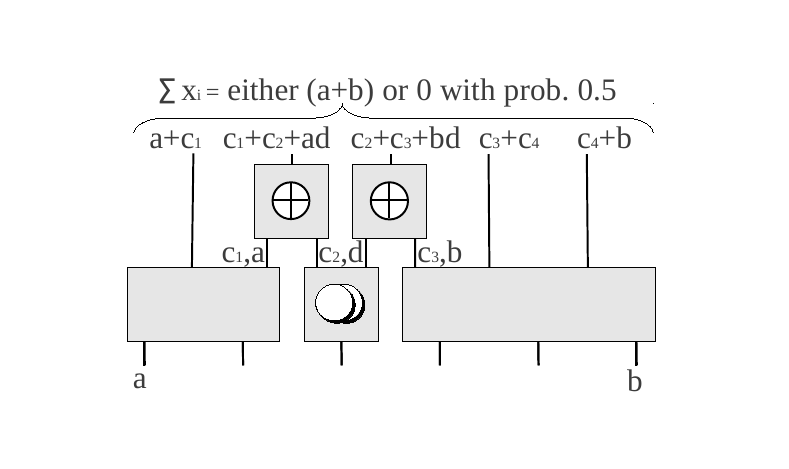}
\caption{$VBox$.\label{fig:VBox} {\small Input cells $1\ldots n-1$ are ignored. Output random variables $(X_i)_{i=1\ldots n}$ have the property that for strict subset $I$ of $1\ldots n$, $X_I$ is uniformly distributed, yet their global parity is, with equal probability, $ab$ or $0$. As such \VBox\ is a localized stochastic map, but can be modified into a translation-invariant stochastic map over entire configurations using the method described for \Parity. \VBox\ is non-signalling, non-correlating and $V$-causal, but it must not be considered a valid PCA: it is not $VV$-causal. The $VBox$ can be seen as a chain of two $GenNLBoxes$}, chaining more of them yields a $V^kBox$.}
\end{figure}

From the \GenNLBox\ example of the previous Section, we are obliged to conclude that a robust notion of the causality of a dynamics in a probabilistic setting cannot be phrased just in terms of a non-signalling or a non-correlation property. Yet, this example has a virtue: it points towards similar questions raised in theoretical physics. Hence, this suggests looking at how those issues were addressed so far in theoretical physics.\\
Indeed, the \NLBox\  is generally regarded as `unphysical' because it does not comply with Bell's \cite{Bell} `local causality', meaning that there is no prior shared resource that one may distribute between Alice and Bob (the outputs of the middle box of Figure \ref{fig:AllNLBox} \emph{(b)}), that will then allow them to perform the required task separately. Distributing quantum resources instead of classical resources (imagining that the outputs of the middle box can now be entangled quantum states) will not fix the problem: yes it does assist Alice and Bob in performing the required task separately \cite{Aspect}; but only approximately so \cite{Tsirelson}.\\
Bell's `local causality' \cite{Bell}, `Screening-off' \cite{HensonSorkin}, `Stochastic Einstein locality' \cite{Hellman,Butterfield}, are all similar conditions, which stem from improvements upon Reichenbach's `Principle of Common Cause'\cite{Szabo,Redei}, as was nicely reviewed in \cite{Henson}. The common cause principle can be summarized as follows: ``Two events can be correlated at a certain time if and only if, conditional upon the events of their mutual past, they become independent''. In the context of this paper, this gives:
\begin{Def}[Screening-off] A stochastic map over configurations $G$ obeys the screening-off condition if and only if for any input cell $i$ with values in $\Sigma$, we have that there exists stochastic maps $(A_x, B_x)_{x\in\Sigma}$ acting over input cells $-\infty\ldots i-1$ and $i+1\ldots +\infty$ respectively, such that for any $\rho$:
$$\rho_i=\tilde{x} \quad\Rightarrow\quad G\rho = (A_x\otimes B_x)\rho.$$
Here input $i$ is said to screen-off $G$.
\end{Def}
This screening-off condition is physically motivated and does not suffer the problems that the non-signalling and non-correlation conditions had. Unfortunately however, it suffers most of the problems of the original, Standard-PCA definition: it is again incomplete and non-composable. For example, \Parity\ does not obey the screening-off condition. Yet, \Parity\ is a natural PCA, clearly implementable by local mechanisms as was shown in Figure \ref{fig:parity}.  But the prior shared resource which is necessary in order to separate Alice from Bob is not present in the input cells, rather it is generated on-the-fly within one time-step of $G$, \cf the circle-marked boxes. In other words, the reason why the screening-off condition rejects \Parity, is because the condition is too stringent in demanding that screening-off events be made explicit in the inputs $x$. A more relaxed condition would be to require that $x$ may be completed so as to then screen-off $G$.
\begin{Def}[Screening-off-completable] A stochastic map over configurations $G$ is screening-off-completable, or simply $V$-causal if and only if for any input cell $i$, we have that there exists stochastic maps (with input/output ranges marked as sub/superscript indices) $A^{-\infty\ldots i-1,l'}_{-\infty\ldots i-1}$, $L^i_{l',l}$, $C^{l,r}_i$, $R^{i+1}_{r,r'}$, $A^{r',i+2\ldots\infty}_{i+1\ldots\infty}$ such that: 
$$G=(L\otimes R)(A\otimes C\otimes B).$$
Here $C$ is said to screen-off $G$ at $i$. See Figure \ref{fig:VCausal} \emph{(a)}.
\end{Def}
Again, is it reasonable to say that PCA are the screening-off-completable, shift-invariant stochastic maps? Again, this is not the case. Indeed, consider a small variation of \Parity, which we call \VBox\ and define in Figure \ref{fig:VBox}. Such a stochastic map is not implementable by local mechanisms: we need some amount of communication between Alice and Bob in order to make it happen. Yet \VBox\ is perfectly $V$-causal, as shown in Figure \ref{fig:VBox}. Hence, whilst the set-up is screening-off-completable, we still must forbid it, our condition is again too weak. A natural family of conditions to consider is $VV$-causality (as defined in Figure \ref{fig:VCausal} \emph{(b)}), $V^3$-causality \etc This route seems a dead end; we believe that the \VBox\ is the $k=1$ instance of the following more general result:
\begin{Con}
For all $k$, there exists a $V^kBox$ which is $V^k$-causal but not $V^{k+1}$-causal.
\end{Con}
\section{Concluding definition}\label{sec:conclusion}

\begin{figure}[h]
\centering
\includegraphics[scale=0.95, clip=true, trim=0.0cm 0.1cm 0.1cm 0.0cm]{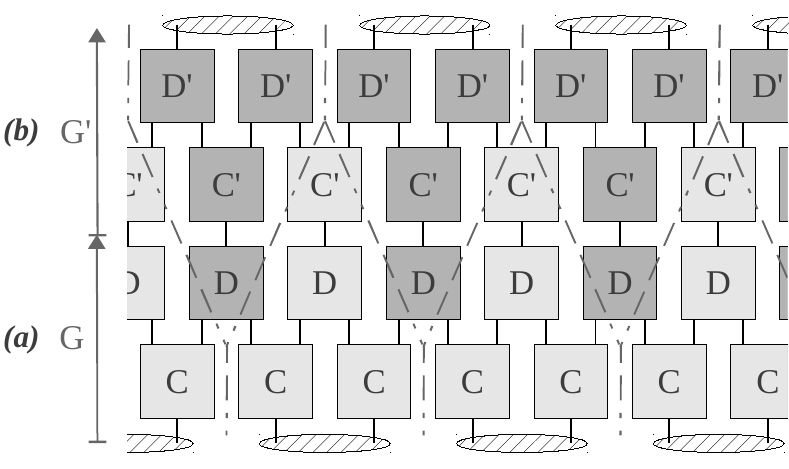}
\caption{
{\bf \emph{(a)}} An operational definition of PCA.\label{fig:PPCA} 
{\small  Time flows upwards. Wires stand for cells. The first two layers of boxes describe one application of a PCA $G=(\bigotimes D)(\bigotimes C)$.}
{\bf \emph{(b)}} Its compositionality. {\small The second two layers describe one application of a PCA $G'=(\bigotimes D')(\bigotimes C')$. The resulting composition is a PCA $G'G=\big(\bigotimes (D'\otimes D')C'D\big)\,\big(\bigotimes C'D(C\otimes C)\big)$ over the hatched supercells of size $2$.}}
\end{figure}

{\em Our best definition.} We have examined several, well-motivated causality principles (non-signalling, non-correlation, $V$-causality) and shown, through a series of surprising counter-examples, that stochastic maps with these properties are not necessarily implementable by local mechanisms. In the limit when $k$ goes to infinity, $V^k$-causality turns into the following definition (assuming shift-invariance):
\begin{Def}[Probabilistic Cellular Automata]\label{def:PPCA}
A stochastic map over configurations $G$ is a PCA if and only if  
$$G=(\bigotimes D)(\bigotimes C)$$
where the $i^{th}$ stochastic matrix $C$ has input $i$ and outputs $l_i,r_i$, 
and the $i^{th}$ stochastic matrix $D$ has inputs $r_{i-1},l_i$ and output $i$, see Figure \ref{fig:PPCA}.  
\end{Def}
This definition has several advantages over Standard-PCA. First, it is complete, because it captures exactly what is meant (up to grouping) by a shift-invariant stochastic map implementable by local mechanisms. Second, it is composable, as shown in Figure \ref{fig:PPCA}. Third, it is less ad-hoc. Indeed, we have mentioned in the introduction that much mathematics is done by having axiomatic definitions and operational description to coincide (usually by means of structure versus characterization theorems). It could be said that the same effect has been achieved in this paper, but in a different manner. Indeed, in this paper we have considered the natural candidate axiomatic definitions of PCA, whose limit is the natural operational description of PCA, and we have discarded each of them  --- by means of counter-examples. In this sense, we have pushed this family of candidate axiomatic definitions to its limit, i.e. as far as to coincide with the operational description.\\
\noindent {\em Discussion.} Still, we cannot really pretend to have reached an axiomatization, nor a characterization of PCA: Definition \ref{def:PPCA} can be presented as just the square of two Standard-PCA; or even more simply as a Standard-PCA with its two phases reversed: first a model of noise gets applied (a stochastic matrix is applied homogeneously on each individual cell), and second, a classical CA is applied. If anything, this paper has shown that it is rather unlikely that an axiomatization can be achieved. The authors are not in the habit of publishing negative results. However, the question of an characterization of PCA \`a la Hedlund has been a long-standing issue: we suspect that many researchers have attempted to obtain such a result in vain. At some point, it becomes just as important to discard the possibility of a Theorem as to establish a new one. Moreover, advances on the issue of causality principles \cite{Bell,Butterfield,Hellman,Szabo,Redei}, have manly arisen from the discovery of counter-examples (See \cite{Henson} and the more recent \cite{CoeckeCausality}), to which this paper adds a number. An impressive amount of literature \cite{PopescuRohrlich,BHK} focusses on the \NLBox\ counter-example (as regards its comparison with quantum information processing, but also its own information processing power) and raise the question of $n$-party extensions: the \GenNLBox\ , the $VBox$, and the $V^kBoxes$ could prove useful in this respect.\\
Finally, let us point out that the observation that the operational description of PCA seems to admit no other axiomatization than itself is in sharp contrast with both the classical case (See Hedlund's theorem Section 1), the reversible case (See \cite{KariCircuit}) and the quantum case (See \cite{ArrighiUCAUSAL}); for which the non-signalling condition alone suffices to entail localizability, \ie implementation by local mechanisms. It is the moment when probabilities come into the picture (whether via stochastic maps or via quantum operations --- this paper cancels out all efforts to generalize the result of \cite{ArrighiUCAUSAL} to an open systems setting) that the non-signalling condition becomes too weak. Moreover, we have seen that replacement principles based on a `common cause' are hardly satisfactory. In philosophy of science it has been argued that the very notion of `physical law' requires an underlying notion of causality, such as non-signalling. But probabilistic/stochastic theories seem to require a very explicit notion of causality, such as localizability.

\section*{Acknowledgements}
The authors would like to acknowledge enlightening discussions with Joe Henson, Jean Mairesse, Jacques Mazoyer, Nicolas Schabanel, Rafa\"el Sorkin and Reinhard Werner.

{\tiny
\bibliographystyle{plain}
\bibliography{../Bibliography/biblio}
}
\end{document}